\pdfoutput=1
\documentclass[12pt,a4paper]{article}

\usepackage{ifthen} 
\newboolean{pdflatex}
\setboolean{pdflatex}{true} 

\newboolean{articletitles}
\setboolean{articletitles}{true} 

\newboolean{uprightparticles}
\setboolean{uprightparticles}{false} 

\def\paperauthors{LHCb collaboration} 
\def\paperasciititle{Observation of the mass difference between neutral charm-meson eigenstates} 
\def\papertitle{Observation of the mass difference between neutral charm-meson eigenstates} 
\def\paperkeywords{{High Energy Physics}, {LHCb}} 
\def\papercopyright{\the\year\ CERN for the benefit of the LHCb collaboration} 
\def\paperlicence{CC BY 4.0 licence}
\def\paperlicenceurl{https://creativecommons.org/licenses/by/4.0/}

\newcommand{\PromptDecay}{\ensuremath{\Dstarp\to\Dz(\to\KS\pip\pim)\pip}\xspace}

\newcommand{\adetpipi} {{\ensuremath{A_{\mathrm{det}}(\pip\pim)}}\xspace}
\newcommand{\adetpipib} {{\ensuremath{A_{\mathrm{det}}^{b}(\pip\pim)}}\xspace}
\newcommand{\adetpipinb} {{\ensuremath{A_{\mathrm{det}}^{-b}(\pip\pim)}}\xspace}

\newcommand{\Ameas}{\ensuremath{A_{\rm meas}}\xspace}

\newcommand{\Dkspp}{\ensuremath{\Dz\to\KS\pip\pim}\xspace}
\newcommand{\Dsphipi}{\ensuremath{\Ds\to\phi\pip}\xspace}
\newcommand{\Dspipipi}{\ensuremath{\Ds\to\pip\pip\pim}\xspace}

\newcommand{\zcp}{\ensuremath{z_{\CP}}\xspace}
\newcommand{\xcp}{\ensuremath{x_{\CP}}\xspace}
\newcommand{\ycp}{\ensuremath{y_{\CP}}\xspace}
\newcommand{\deltaz}{\ensuremath{\Delta z}\xspace}
\newcommand{\deltax}{\ensuremath{\Delta x}\xspace}
\newcommand{\deltay}{\ensuremath{\Delta y}\xspace}

\newcommand{\re}[2][()]{\ifthenelse{\equal{#1}{()}}{{\ensuremath{{\rm \, Re}}\!\left(#2\right)}}{{\ensuremath{{\rm \, Re}}\!\left[#2\right]}}}
\newcommand{\im}[2][()]{\ifthenelse{\equal{#1}{()}}{{\ensuremath{{\rm \, Im}}\!\left(#2\right)}}{{\ensuremath{{\rm \, Im}}\!\left[#2\right]}}}

\newcommand{\xcpStat}{\ensuremath{0.46}}
\newcommand{\ycpStat}{\ensuremath{1.20}}
\newcommand{\dxStat}{\ensuremath{0.18}}
\newcommand{\dyStat}{\ensuremath{0.36}}

\newcommand{\xcpRes}{\ensuremath{3.97}}
\newcommand{\ycpRes}{\ensuremath{4.59}}
\newcommand{\dxRes}{\ensuremath{-0.27}}
\newcommand{\dyRes}{\ensuremath{0.20}}

\newcommand{\xcpSyst}{\ensuremath{0.29}}
\newcommand{\ycpSyst}{\ensuremath{0.85}}
\newcommand{\dxSyst}{\ensuremath{0.01}}
\newcommand{\dySyst}{\ensuremath{0.13}}

\newcommand{\xUnits}{\times10^{-3}}
\newcommand{\yUnits}{\times10^{-3}}
\newcommand{\dxUnits}{\times10^{-3}}
\newcommand{\dyUnits}{\times10^{-3}}

\usepackage[top=1in, bottom=1.25in, left=1in, right=1in]{geometry}

\usepackage{microtype}
\usepackage{lineno}  
\usepackage{xspace} 
\usepackage{caption} 

\usepackage{graphicx}  
\usepackage{color}
\usepackage{colortbl}
\graphicspath{{./figs/}} 

\usepackage{amsmath} 
\usepackage{amssymb}
\usepackage{amsfonts}
\usepackage{upgreek} 

\newcommand*\patchAmsMathEnvironmentForLineno[1]{%
\expandafter\let\csname old#1\expandafter\endcsname\csname #1\endcsname
\expandafter\let\csname oldend#1\expandafter\endcsname\csname
end#1\endcsname
 \renewenvironment{#1}%
   {\linenomath\csname old#1\endcsname}%
   {\csname oldend#1\endcsname\endlinenomath}%
}
\newcommand*\patchBothAmsMathEnvironmentsForLineno[1]{%
  \patchAmsMathEnvironmentForLineno{#1}%
  \patchAmsMathEnvironmentForLineno{#1*}%
}
\AtBeginDocument{%
\patchBothAmsMathEnvironmentsForLineno{equation}%
\patchBothAmsMathEnvironmentsForLineno{align}%
\patchBothAmsMathEnvironmentsForLineno{flalign}%
\patchBothAmsMathEnvironmentsForLineno{alignat}%
\patchBothAmsMathEnvironmentsForLineno{gather}%
\patchBothAmsMathEnvironmentsForLineno{multline}%
\patchBothAmsMathEnvironmentsForLineno{eqnarray}%
}

\usepackage{hyperxmp}

\usepackage[pdftex,
            pdfauthor={\paperauthors},
            pdftitle={\paperasciititle},
            pdfkeywords={\paperkeywords},
            pdfcopyright={Copyright (C) \papercopyright},
            pdflicenseurl={\paperlicenceurl}]{hyperref}

\usepackage[colorinlistoftodos,textsize=scriptsize]{todonotes}

\usepackage[bottom,flushmargin,hang,multiple]{footmisc}

\usepackage[all]{hypcap} 

\usepackage{xspace} 
\usepackage{upgreek}

\def\lhcb   {\mbox{LHCb}\xspace}

\def\velo   {VELO\xspace}

\def\MagUp {\mbox{\em Mag\kern -0.05em Up}\xspace}

\ifthenelse{\boolean{uprightparticles}}%
{

 \def\Ppi         {\ensuremath{\uppi}\xspace}

 \def\PDelta      {\ensuremath{\Delta}\xspace}                 
 \def\PXi         {\ensuremath{\Xi}\xspace}                 
 \def\PLambda     {\ensuremath{\Lambda}\xspace}                 
 \def\PSigma      {\ensuremath{\Sigma}\xspace}                 
 \def\POmega      {\ensuremath{\Omega}\xspace}                 
 \def\PUpsilon    {\ensuremath{\Upsilon}\xspace}

 \def\PB      {\ensuremath{\mathrm{B}}\xspace}                 
                  
 \def\PD      {\ensuremath{\mathrm{D}}\xspace}

 \def\PK      {\ensuremath{\mathrm{K}}\xspace}

 \def\Pb      {\ensuremath{\mathrm{b}}\xspace}                 
 \def\Pc      {\ensuremath{\mathrm{c}}\xspace}

 \def\Pi      {\ensuremath{\mathrm{i}}\xspace}

 \def\Ps      {\ensuremath{\mathrm{s}}\xspace}

 \def\thebaroffset{0.0em}
}
{

 \def\Ppi         {\ensuremath{\pi}\xspace}

 \mathchardef\PDelta="7101
 \mathchardef\PXi="7104
 \mathchardef\PLambda="7103
 \mathchardef\PSigma="7106
 \mathchardef\POmega="710A
 \mathchardef\PUpsilon="7107
                  
 \def\PB      {\ensuremath{B}\xspace}                 
                  
 \def\PD      {\ensuremath{D}\xspace}

 \def\PK      {\ensuremath{K}\xspace}

 \def\Pb      {\ensuremath{b}\xspace}                 
 \def\Pc      {\ensuremath{c}\xspace}

 \def\Pi      {\ensuremath{i}\xspace}

 \def\Ps      {\ensuremath{s}\xspace}

 \def\thebaroffset{0.18em}
}
\newcommand{\offsetoverline}[2][\thebaroffset]{\kern #1\overline{\kern -#1 #2}}%

\makeatletter
\ifcase \@ptsize \relax
  \newcommand{\miniscule}{\@setfontsize\miniscule{4}{5}}
\or
  \newcommand{\miniscule}{\@setfontsize\miniscule{5}{6}}
\or
  \newcommand{\miniscule}{\@setfontsize\miniscule{5}{6}}
\fi
\makeatother

\DeclareRobustCommand{\optbar}[1]{\shortstack{{\miniscule (\rule[.5ex]{1.25em}{.18mm})}
  \\ [-.7ex] $#1$}}

\def\squark    {{\ensuremath{\Ps}}\xspace}

\def\cquark    {{\ensuremath{\Pc}}\xspace}

\def\bquark    {{\ensuremath{\Pb}}\xspace}

\def\pion   {{\ensuremath{\Ppi}}\xspace}

\def\pip    {{\ensuremath{\pion^+}}\xspace}
\def\pim    {{\ensuremath{\pion^-}}\xspace}

\def\kaon    {{\ensuremath{\PK}}\xspace}

\def\KorKbar {\kern \thebaroffset\optbar{\kern -\thebaroffset \PK}{}\xspace}

\def\Kp      {{\ensuremath{\kaon^+}}\xspace}
\def\Km      {{\ensuremath{\kaon^-}}\xspace}

\def\KS      {{\ensuremath{\kaon^0_{\mathrm{S}}}}\xspace}

\def\Kstar   {{\ensuremath{\kaon^*}}\xspace}

\def\Dbar    {{\ensuremath{\offsetoverline{\PD}}}\xspace}
\def\D       {{\ensuremath{\PD}}\xspace}

\def\DorDbar {\kern \thebaroffset\optbar{\kern -\thebaroffset \PD}\xspace}
\def\Dz      {{\ensuremath{\D^0}}\xspace}
\def\Dzb     {{\ensuremath{\Dbar{}^0}}\xspace}
\def\Dp      {{\ensuremath{\D^+}}\xspace}
\def\Dm      {{\ensuremath{\D^-}}\xspace}

\def\DpDm    {\ensuremath{\Dp {\kern -0.16em \Dm}}\xspace}
\def\Dstar   {{\ensuremath{\D^*}}\xspace}

\def\Dstarp  {{\ensuremath{\D^{*+}}}\xspace}

\def\Ds      {{\ensuremath{\D^+_\squark}}\xspace}
\def\Dsp     {{\ensuremath{\D^+_\squark}}\xspace}
\def\Dsm     {{\ensuremath{\D^-_\squark}}\xspace}

\def\B       {{\ensuremath{\PB}}\xspace}

\def\BorBbar {\kern \thebaroffset\optbar{\kern -\thebaroffset \PB}\xspace}

\def\Bd      {{\ensuremath{\B^0}}\xspace}

\def\BdorBdbar {\kern \thebaroffset\optbar{\kern -\thebaroffset \Bd}\xspace}

\def\Bs      {{\ensuremath{\B^0_\squark}}\xspace}

\def\BsorBsbar {\kern \thebaroffset\optbar{\kern -\thebaroffset \Bs}\xspace}

\def\Y#1S{\ensuremath{\PUpsilon{(#1S)}}\xspace}

\def\LorLbar     {\kern \thebaroffset\optbar{\kern -\thebaroffset \PLambda}\xspace}

\newcommand{\decay}[2]{\ensuremath{#1\!\to #2}\xspace} 

\def\to                 {\ensuremath{\rightarrow}\xspace}

\def\CP                {{\ensuremath{C\!P}}\xspace}

\newcommand{\dm}{{\ensuremath{\Delta m}}\xspace}

\def\AT#1     {\ensuremath{A_{\mathrm{T}}^{#1}}\xspace}           

\def\C#1      {\ensuremath{\mathcal{C}_{#1}}\xspace}                       
\def\Cp#1     {\ensuremath{\mathcal{C}_{#1}^{'}}\xspace}                    
\def\Ceff#1   {\ensuremath{\mathcal{C}_{#1}^{\mathrm{(eff)}}}\xspace}        
\def\Cpeff#1  {\ensuremath{\mathcal{C}_{#1}^{'\mathrm{(eff)}}}\xspace}       
\def\Ope#1    {\ensuremath{\mathcal{O}_{#1}}\xspace}                       
\def\Opep#1   {\ensuremath{\mathcal{O}_{#1}^{'}}\xspace}                    

\def\ycp        {\ensuremath{y_{\CP}}\xspace}

\newcommand{\aunit}[1]{\ensuremath{\text{\,#1}}}       

\newcommand{\tev}{\aunit{Te\kern -0.1em V}\xspace}
\newcommand{\gev}{\aunit{Ge\kern -0.1em V}\xspace}
\newcommand{\mev}{\aunit{Me\kern -0.1em V}\xspace}
\newcommand{\kev}{\aunit{ke\kern -0.1em V}\xspace}
\newcommand{\ev}{\aunit{e\kern -0.1em V}\xspace}
 
\newcommand{\mevc}{\ensuremath{\aunit{Me\kern -0.1em V\!/}c}\xspace}
\newcommand{\gevc}{\ensuremath{\aunit{Ge\kern -0.1em V\!/}c}\xspace}
\newcommand{\mevcc}{\ensuremath{\aunit{Me\kern -0.1em V\!/}c^2}\xspace}
\newcommand{\gevcc}{\ensuremath{\aunit{Ge\kern -0.1em V\!/}c^2}\xspace}
\newcommand{\gevgevcccc}{\ensuremath{\gev^2\!/c^4}\xspace} 

\def\fb   {\ensuremath{\aunit{fb}}\xspace}
\def\invfb   {\ensuremath{\fb^{-1}}\xspace}

\def\fs   {\aunit{fs}}

\newcommand{\stat}{\aunit{(stat)}\xspace}
\newcommand{\syst}{\aunit{(syst)}\xspace}

\newcommand{\chisq}{\ensuremath{\chi^2}\xspace}

\def\gsim{{~\raise.15em\hbox{$>$}\kern-.85em
          \lower.35em\hbox{$\sim$}~}\xspace}
\def\lsim{{~\raise.15em\hbox{$<$}\kern-.85em
          \lower.35em\hbox{$\sim$}~}\xspace}

\def\tell1  {TELL1\xspace}
\def\ukl1   {UKL1\xspace}

\usepackage{cite} 
\usepackage{mciteplus}

\usepackage{longtable} 

\graphicspath{{./figs/}}
\usepackage[capitalise]{cleveref}
\usepackage{booktabs}
\Crefname{figure}{Fig.}{Figs.}

\begin{document}

\renewcommand{\thefootnote}{\fnsymbol{footnote}}
\setcounter{footnote}{1}


\begin{titlepage}
\pagenumbering{roman}

\vspace*{-1.5cm}
\centerline{\large EUROPEAN ORGANIZATION FOR NUCLEAR RESEARCH (CERN)}
\vspace*{1.5cm}
\noindent
\begin{tabular*}{\linewidth}{lc@{\extracolsep{\fill}}r@{\extracolsep{0pt}}}
\ifthenelse{\boolean{pdflatex}}
{\vspace*{-1.5cm}\mbox{\!\!\!\includegraphics[width=.14\textwidth]{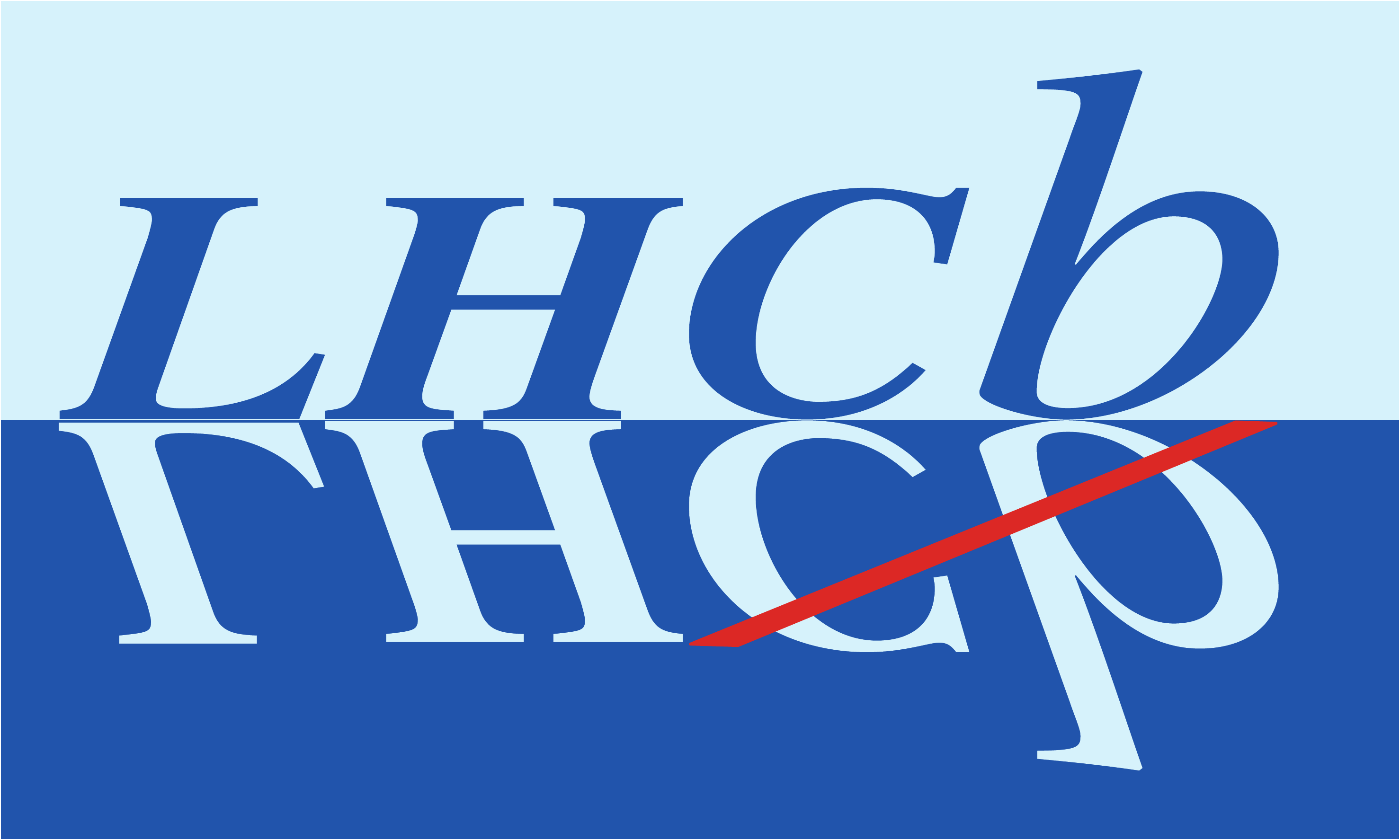}} & &}%
{\vspace*{-1.2cm}\mbox{\!\!\!\includegraphics[width=.12\textwidth]{figs/lhcb-logo.eps}} & &}%
\\
 & & CERN-EP-2021-099 \\  
 & & LHCb-PAPER-2021-009 \\  
 & & August 17, 2023 \\ 
 & & \\
\end{tabular*}

\vspace*{3.0cm}

{\normalfont\bfseries\boldmath\huge
\begin{center}
  \papertitle 
\end{center}
}

\vspace*{1.5cm}

\begin{center}
\paperauthors\footnote{Authors are listed at the end of this paper.}
\end{center}

\vspace{\fill}

\begin{abstract}
  \noindent

  A measurement of mixing and \CP violation in neutral charm mesons is performed using data reconstructed in proton--proton collisions collected by the \lhcb experiment from 2016 to 2018, corresponding to an integrated luminosity of 5.4\invfb. A total of $30.6$ million \Dkspp decays are analyzed using a method optimized for the measurement of the mass difference between neutral charm-meson eigenstates. Allowing for \CP violation in mixing and in the interference between mixing and decay, the mass and decay-width differences are measured to be $\xcp = \left[\xcpRes\pm\xcpStat\stat\pm\xcpSyst\syst\right]\xUnits$ and $\ycp = \left[\ycpRes\pm\ycpStat\stat\pm\ycpSyst\syst\right]\yUnits$, respectively. The \CP-violating parameters are measured as $\deltax = \left[\dxRes\pm\dxStat\stat\pm\dxSyst\syst\right]\dxUnits$ and $\deltay = \left[\dyRes\pm\dyStat\stat\pm\dySyst\syst\right]\dyUnits$. This is the first observation of a nonzero mass difference in the \Dz meson system, with a significance exceeding seven standard deviations. The data are consistent with \CP symmetry, and improve existing constraints on the associated parameters.

\end{abstract}

\vspace*{1.5cm}
\begin{center}
  Published in
  Phys.~Rev.~Lett. \textbf{127} (2021) 111801
\end{center}

\vspace{\fill}

{\footnotesize 
\centerline{\copyright~\papercopyright. \href{\paperlicenceurl}{\paperlicence}.}}
\vspace*{2mm}

\end{titlepage}


\newpage
\setcounter{page}{2}
\mbox{~}
%
%
%
%


\renewcommand{\thefootnote}{\arabic{footnote}}
\setcounter{footnote}{0}

\cleardoublepage


\pagestyle{plain} 
\setcounter{page}{1}
\pagenumbering{arabic}


Neutral charm mesons propagating freely can change (oscillate) into their own antiparticles, as the mass eigenstates are linear combinations of the flavor eigenstates. These flavor-changing neutral currents do not occur at tree level in the Standard Model (SM) and allow for hypothetical particles of arbitrarily high mass to contribute significantly to the process. This can affect the mixing of mesons and antimesons such that measurements of these processes can probe physics beyond the SM~\cite{Golowich:2007ka,Isidori:2010kg,Bobrowski:2010xg,UTfit:2007eik}.

The mass eigenstates of charm mesons can be written as \mbox{$|D_{1,2}\rangle\equiv p|\Dz\rangle\pm q |\Dzb\rangle$}, where $p$ and $q$ are complex parameters and, in the limit of charge-parity (\CP) symmetry, $|D_{1}\rangle$ ($|D_{2}\rangle$) is defined as the \CP even (odd) eigenstate. Mixing of flavor eigenstates is described by the dimensionless parameters \mbox{$x\equiv(m_1 - m_2)c^2/\Gamma$} and \mbox{$y\equiv(\Gamma_1 - \Gamma_2)/(2\Gamma)$}, where $m_{1(2)}$ and $\Gamma_{1(2)}$ are the mass and decay width of the $D_{1(2)}$ state, respectively, and $\Gamma$ 
is the average decay width~\cite{PDG2020}.
In \Dz and \Dzb decays to a common final state, $f$, \CP violation in mixing manifests itself if $|q/p|\neq1$ or in the interference between mixing and decay if \mbox{$\phi_f\equiv\arg(q\bar{A}_f/pA_f)\neq0$}. Here $A_f$ ($\bar{A}_f$) denotes the amplitude of the decay process $\Dz\to f$ ($\Dzb\to f$). In the \Dkspp decay studied in this Letter, \CP violation in the decay ($|A_f|^2\neq|\bar{A}_f|^2$) is not considered, as in the SM it is negligible for the doubly Cabibbo-suppressed (DCS) and Cabibbo-favored (CF) amplitudes contributing to this process. With this assumption, the \CP-violating phase is independent of the final state, $\phi_f\approx\phi\approx\arg(q/p)$~\cite{Du:1986ai,Bergmann:2000id}.  

The current world average of the mixing and \CP-violating parameters yields \mbox{$x = (3.7\pm1.2)\times10^{-3}$}, $y = (6.8\,^{+\,0.6}_{-\,0.7})\times10^{-3}$, $|q/p|=0.951\,^{+\,0.053}_{-\,0.042}$, and $\phi=-0.092\,^{+\,0.085}_{-\,0.079}$~\cite{HFLAV18}. Measurements using decays such as $\Dz\to\Kp\pim$ have resulted in precise measurements of $y$ and have allowed for the observation of mixing~\cite{Aubert:2007wf,Staric:2007dt}. However, the data remain marginally compatible with $x=0$, and are consistent with \CP symmetry. Theoretical predictions for the mixing parameters are of similar magnitude but less precise~\cite{Bigi:2000wn,Falk:2004wg}, while predictions of the \CP-violating phase are around 0.002 \cite{PhysRevD.103.053008} and are well below the current experimental precision.

\begin{figure}[t]
\centering
\includegraphics[width=0.3\textwidth]{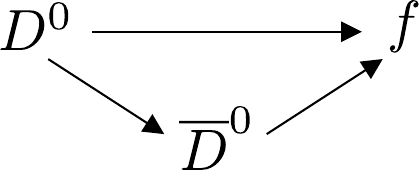}
\caption{Depiction of the interference of mixing and decay if a $\Dz$ and a $\Dzb$ meson decay to a common final state $f$. \label{fig:mixing}}
\end{figure}

Sensitivity to the mixing and \CP-violating parameters is offered by the self-conjugate, multibody \Dkspp decay~\cite{Asner:2005sz,Peng:2014oda,delAmoSanchez:2010xz,LHCb-PAPER-2015-042,LHCb-PAPER-2019-001}. Inclusion of the charge-conjugate process is implied unless stated otherwise. This final state is accessible in both $\Dz$ and $\Dzb$ decays and leads to interference between the mixing and decay amplitudes, as demonstrated pictorially in \cref{fig:mixing}. The dynamics of the decay are expressed as a function of two invariant masses following the Dalitz-plot formalism, in which a three-body decay is parametrized by a pair of two-body invariant masses~\cite{Dalitz:1953cp,Fabri:1954zz}. The squared invariant mass $m^2(\KS\pi^\pm)$ is denoted as $m_\pm^2$ for $\Dz$ decays and $m_\mp^2$ for $\Dzb$ decays. A mixture of DCS and CF decay amplitudes results in large variations of the strong phase and, with mixing, causes a decay-time evolution of the density of decays across the phase space.  A joint analysis of the Dalitz-plot and decay-time distributions may be used to determine the mixing parameters. Splitting the sample by flavor of the charm meson at production probes for \CP-violating effects. Usage of multibody decay modes is typically challenging, as it requires knowledge of the variation of the hadronic parameters and excellent control of efficiencies, resolutions, and background effects.  

This Letter reports on a measurement of the mixing and \CP violation parameters in \Dkspp decays using the ``bin-flip'' method~\cite{binflip-paper}, a model-independent approach which obviates the need for detailed models of the efficiency, resolution, and contributing amplitudes. Mixing and \CP violation are parametrized by \zcp and \deltaz, which are defined by $\zcp\pm\deltaz\equiv-\left(q/p\right)^{\pm1}(y+ix)$. The results are expressed in terms of the \CP-even mixing parameters \mbox{$\xcp\equiv-\im{\zcp}$} and \mbox{$\ycp\equiv-\re{\zcp}$}, and of the \CP-violating differences \mbox{$\deltax\equiv-\im{\deltaz}$} and \mbox{$\deltay\equiv-\re{\deltaz}$}. Conservation of \CP symmetry implies $\xcp=x$, $\ycp=y$, and $\deltax=\deltay=0$. The method has already been employed by the LHCb collaboration, yielding the single most precise measurement of $\xcp$ and \deltax~\cite{LHCb-PAPER-2019-001}. 

\begin{figure}[t]
\centering
\includegraphics[width=0.5\textwidth]{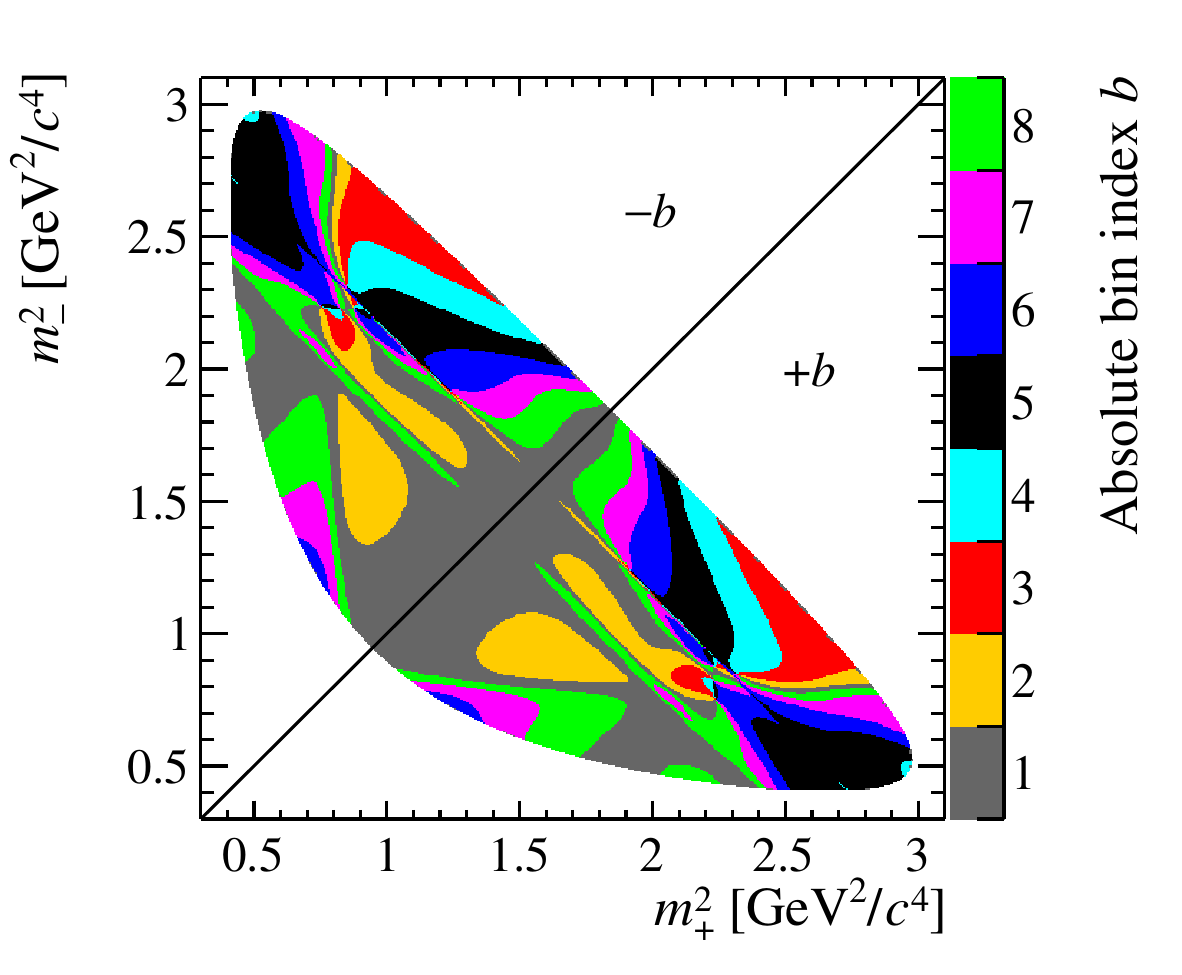}
\caption{``Binning" of the \Dkspp Dalitz plot. Colors indicate the absolute value of the bin index $b$. \label{fig:binning}}
\end{figure}

In the bin-flip method, data are partitioned into disjoint regions (bins) of the Dalitz plot, which are defined to preserve nearly constant strong-phase differences $\Delta\delta(m_-^2,m_+^2)$ between the \Dz and \Dzb amplitudes within each bin~\cite{Libby:2010nu}. Two sets of eight bins are formed symmetrically about the $m_+^2 = m_-^2$ bisector, as illustrated in \cref{fig:binning}. The region satisfying $m_+^2 > m_-^2$, which includes regions dominated by the CF \decay{\Dz}{\Kstar(892)^-\pi^+} decay, is given a positive index $+b$, while the opposite region, where the relative contribution from decays following an oscillation is enhanced, is given a negative index $-b$. The data are further split into 13 bins of decay time, chosen such that the bins are approximately equally populated. The squared-mass and decay-time resolutions are typically $0.006$\gevgevcccc and $60\fs$, respectively, which are smaller than the bin sizes used. Thus, they are neglected and accounted for in the systematic uncertainties. 

For each decay-time interval ($j$), the ratio of the number of decays in each negative Dalitz-plot bin ($-b$) to its positive counterpart ($+b$) is measured. The usage of ratios minimizes the need for precise modeling of the efficiency variation across phase space and decay time.  For small mixing parameters and \CP-conserving decay amplitudes, the expected ratios for initially produced \Dz (\Dzb) mesons, $R_{bj}^+$ ($R_{bj}^-$), are~\cite{binflip-paper}
\begin{equation}\label{eq:bin-flip-ratio}
R_{bj}^\pm \approx \frac{r_b + r_b \dfrac{\langle t^2\rangle_j}{4}\re{\zcp^2-\deltaz^2} + \dfrac{\langle t^2\rangle_j}{4}\left|\zcp\pm\deltaz\right|^2 + \sqrt{r_b}\langle t\rangle_j \re[]{X_b^*(\zcp\pm\deltaz)}}{1 + \dfrac{\langle t^2\rangle_j}{4}\re{\zcp^2-\deltaz^2}+r_b \dfrac{\langle t^2\rangle_j}{4}\left|\zcp\pm\deltaz\right|^2 + \sqrt{r_b} \langle t\rangle_j \re[]{X_b(\zcp\pm\deltaz)}}.
\end{equation}
The parameter $r_b$ is the value of  $R_{bj}$ at $t=0$, while $X_b$ is the amplitude-weighted strong-phase difference between opposing bins. Finally, $\langle t\rangle_j$ ($\langle t^2\rangle_j$) corresponds to the average (squared) decay time in each positive Dalitz-plot region where the mixed contribution is negligible, in units of the \Dz lifetime $\tau=\hbar/\Gamma$~\cite{PDG2020}, calculated directly from background-subtracted data. The other parameters are determined from a simultaneous fit of the observed $R_{bj}^\pm$ ratios, in which external information on $c_b\equiv\re{X_b}$ and \mbox{$s_b\equiv-\im{X_b}$}~\cite{Libby:2010nu,Ablikim:2020lpk} is used as a constraint.

Samples of \Dkspp decays are reconstructed from proton--proton ($pp$) collisions collected by the \lhcb experiment from 2016 to 2018, corresponding to an integrated luminosity of 5.4\invfb. The strong-interaction decay $\Dstarp\to\Dz\pip$ is used to identify the flavor of the neutral charm meson at production. Throughout this Letter, \Dstarp indicates the $\Dstar(2010)^+$ meson and soft pion indicates the pion from its decay. The LHCb detector~\cite{LHCb-DP-2008-001,LHCb-DP-2014-002} is a single-arm forward spectrometer covering the pseudorapidity range $2 < \eta < 5$, designed for the study of particles containing \bquark\ or \cquark\ quarks.

Decays of \decay{\KS}{\pip\pim} are reconstructed in two different categories: the first involving \KS mesons that decay early enough for the pions to be reconstructed in all tracking detectors; and the second containing \KS mesons that decay later such that track segments of the pions cannot be formed in the vertex detector, which surrounds the $pp$ interaction (primary vertex) region, resulting in a worse momentum resolution. The latter category contains more candidates but has slightly worse mass and decay-time resolution as well as larger efficiency variations. 

The online event selection consists of a hardware stage, selecting events based on calorimeter and muon detector information, followed by two software stages. In the first software stage, the pion pair from the $\Dz$ decay is required to satisfy criteria on momenta and final-state charged-particle displacements from any primary vertex for at least one pion (one-track) or both together with a vertex quality requirement (two-track). The second software stage fully reconstructs  \mbox{$\Dstarp\to\Dz\pip,\Dkspp$} candidates using further requirements on particle identification, momenta, and track and vertex quality. Specific ranges of displacement and invariant mass are imposed on the reconstructed $\Dz$ and $\KS$ candidates. Due to differing efficiencies, the sample is split into four categories, depending on whether or not the \KS meson is reconstructed in the \velo and whether or not they satisfy the one-track requirement.

Offline, a kinematic fit constrains the tracks to form vertices according to the decay topology, the \KS candidate mass to the known value~\cite{PDG2020}, and the \Dstarp candidate to a primary vertex~\cite{Hulsbergen:2005pu}. In the reconstruction of the Dalitz-plot coordinates, an additional constraint on the \Dz candidate mass to the known value improves the resolution. Charm mesons originating from the decays of \bquark hadrons are suppressed by requiring that the \Dz and soft pion candidates originate from a primary vertex. Candidates are rejected if two of the reconstructed tracks use the same hits in the vertex detector. About 6\% of the candidates are from collision events in which multiple candidates are reconstructed, usually by pairing the same \Dz candidate with different soft pions. When this occurs, one candidate is chosen randomly, and the rest are removed from the sample.

\begin{figure}[t]
\centering
\includegraphics[width=0.5\textwidth]{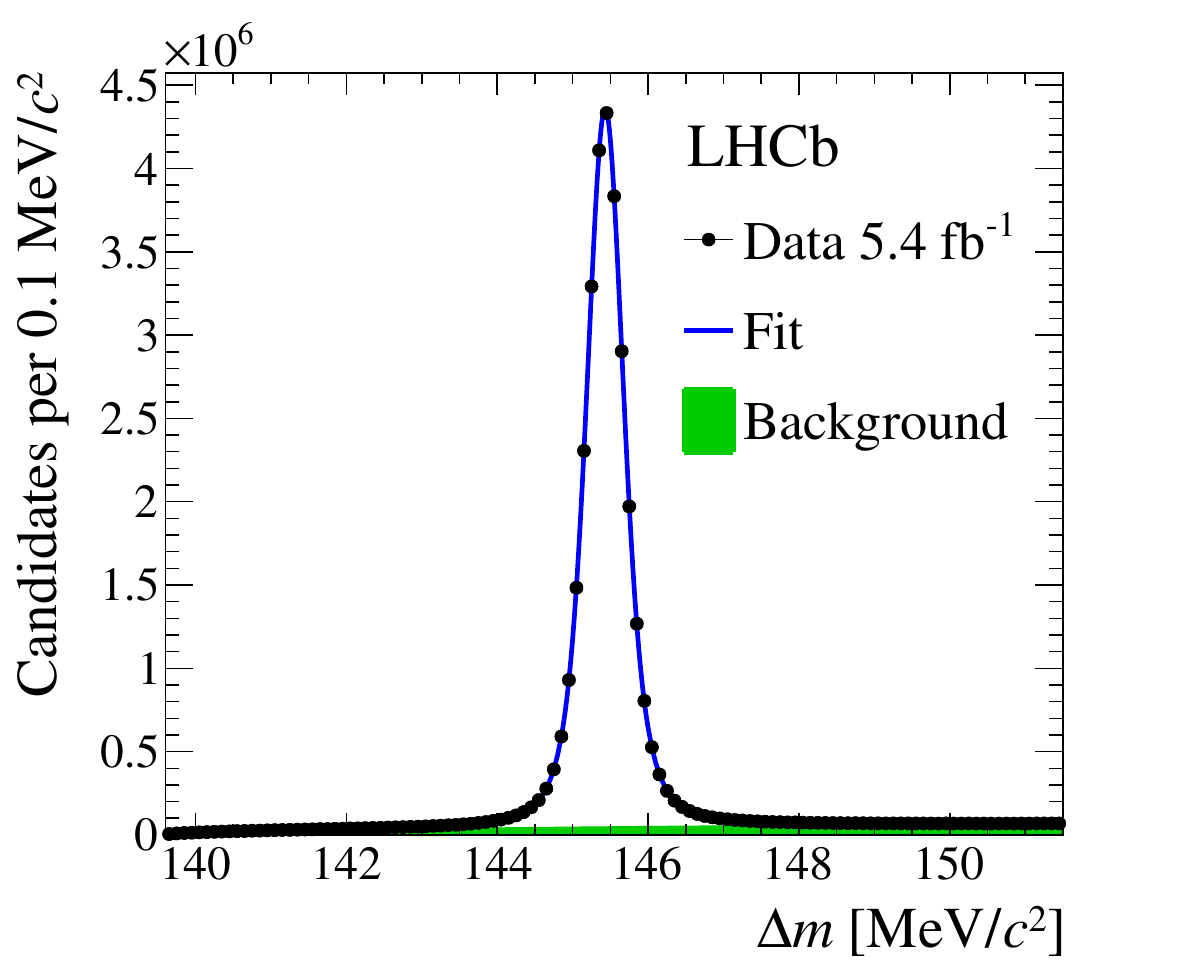}\hfil
\caption{ Distribution of $\dm$ for the selected \PromptDecay candidates. The projection of the fit result is superimposed.
\label{fig:example-fits}}
\end{figure}

Signal yields are determined by fitting the distribution of the mass difference between the \Dstarp and \Dz candidates, denoted as \dm. The signal probability density function is empirically described by a combination of a Johnson $S_U$ distribution~\cite{johnson} and two Gaussian functions, one of which shares a mean with the Johnson $S_U$. The background is dominated by real \Dz decays incorrectly combined with a charged particle not associated with a \Dstarp decay, and is modeled with a smooth phase-space-like model,   \mbox{$\theta(\dm-m_{\pi})e^{-c(\dm-m_{\pi})}\left(\dm-m_{\pi}\right)^{\alpha}$}, where $\theta(x)$ is the Heaviside step function, $m_{\pi}$ is the charged-pion mass~\cite{PDG2020}, and $\alpha$ and $c$ are free parameters. Figure 3 shows the \dm distribution of the entire sample, from which the fit identifies $(30.585\pm 0.011) \times 10^6$ signal decays. This represents a factor of 15 larger yield compared to the previous measurement.

To determine the yields used to form the ratios $R_{bj}^{\pm}$, separate fits are performed for each set of Dalitz-plot and decay-time bins $bj$. The signal model assumes the same parameters for each pair of positive and negative Dalitz-plot bins, and fixes some parameters from a fit integrated over decay time. Fits are performed independently for \Dz and \Dzb candidates, as well as for each of the four data subsamples. The measured signal yields are then corrected for two effects that do not cancel in the ratio: experimentally induced correlations between the phase space and decay time, and charge-dependent efficiencies (detection asymmetries).

Online requirements on the displacement and momenta of the \Dz decay products introduce efficiency variations that are correlated between the phase-space coordinates and the \Dz decay time. The effect depends predominantly on the invariant mass of two pions from the \Dz decay, with the efficiency to reconstruct the candidates at low values decreasing significantly at low \Dz decay times. This can bias the measured yield ratios and produce mixing-like trends. To remove this bias, an approach that estimates the relative efficiencies using data is developed. The Dalitz plot is divided into small, rectangular-like regions formed symmetrically across the bisector. Note that these include the portions above and below the bisector, unlike the bins shown in \cref{fig:binning}. In the limit of \CP symmetry, the contribution of mixing to such symmetric regions depends only on $\ycp$ and the hadronic parameters of the $\Dz$ decay~\cite{binflip-paper}. As oscillations result in a migration of decays from one side of the Dalitz plot to the other, and the regions are symmetric with respect to the bisector, there is no effect from $\xcp$. Given a set of inputs for $\ycp$ and the hadronic parameters, the contribution of mixing to the decay-time distributions of these regions can be accounted for, such that the remaining differences between regions come from the efficiency correlations. Relative efficiency maps that align the decay-time distributions in all these symmetric regions can then be calculated. Per-candidate weights assigned by the efficiency maps are integrated over the data in each bin using the sPlot method~\cite{Pivk:2004ty} with \dm as the discriminating variable. This provides correction factors for each of the fitted signal yields. 

In calculating the efficiency maps, the strong phase variation within a Dalitz-plot bin is approximated as constant, such that it can be described by the external inputs ($s_b$). As $\ycp$ and $s_b$ are parameters of the fit, the correction maps and corresponding correction factors are calculated for a range of values. The smallness of mixing results in smooth variations of the correction factors for a given Dalitz-plot bin, which allows for precise interpolation between the calculated points with polynomials. These polynomials are then incorporated into the fit as a correction that depends on $\ycp$ and $s_b$. The correction is calculated for each yield ratio, but is averaged over the initial flavor of the candidates. The procedure has been validated with pseudoexperiments, and a systematic uncertainty is assigned due to the approximation that $s_b$ is constant within a bin.

Corrections are also applied in order to take into account detection asymmetries. Due to utilizing ratios of yields, the analysis is insensitive to detection asymmetries of the \KS, as well as the soft pion used to tag the flavor of the candidate. However, the kinematics of the pions produced in the \Dz decay depend on the Dalitz-plot coordinate and \Dz flavor. This can result in asymmetric efficiency variations for \Dz and \Dzb candidates that imitate \CP violation. The two-track $\pip\pim$ asymmetry, $\adetpipi$, is determined by measuring detection asymmetries in control samples of $\Dspipipi$ and $\Dsphipi$ decays, in which the $\phi$ meson is reconstructed through a \Kp\Km pair. A randomly chosen  $\pip$ in the $\Dspipipi$ decay is paired with the \pim to form a proxy for the $\pip\pim$ pair of interest. The $\Dsphipi$ sample is used to cancel asymmetries induced from the remaining $\pip$, $A_{\text{det}}(\pip)$, and other sources, such as the trigger selection, $A_{\text{trigger}}(\Ds)$, and the production of $\Dsp$ and $\Dsm$ mesons in $pp$ collisions, $A_{\text{prod}}(\Ds)$. For asymmetries of $\mathcal{O}(1\%)$, the raw asymmetries $\Ameas$ can be approximated as
\begin{equation}
\begin{aligned}
    \Ameas(\Dspipipi) &\approx \adetpipi\ +&A_{\text{det}}(\pip) + A_{\text{prod}}(\Ds) + A_{\text{trigger}}(\Ds),\\
    \Ameas(\Dsphipi) &\approx &A_{\text{det}}(\pip) + A_{\text{prod}}(\Ds) + A_{\text{trigger}}(\Ds).
\end{aligned}
\end{equation}

The difference of the two measured asymmetries gives the detection asymmetry of the \pip\pim pair. The control samples are weighted to match the kinematics of the pions from the \Dkspp sample. This weighting is done separately for each Dalitz-plot bin. The detection asymmetries are of the order of $10^{-3}$ and are used as corrections to the measured yields.  They are included as constraints in the fit along with the associated covariance matrix $\Delta V_{\rm{asym}}$ describing uncertainties coming from the limited size of the calibration samples.

\begin{figure}[p]
\centering
\includegraphics[width=0.7\textwidth]{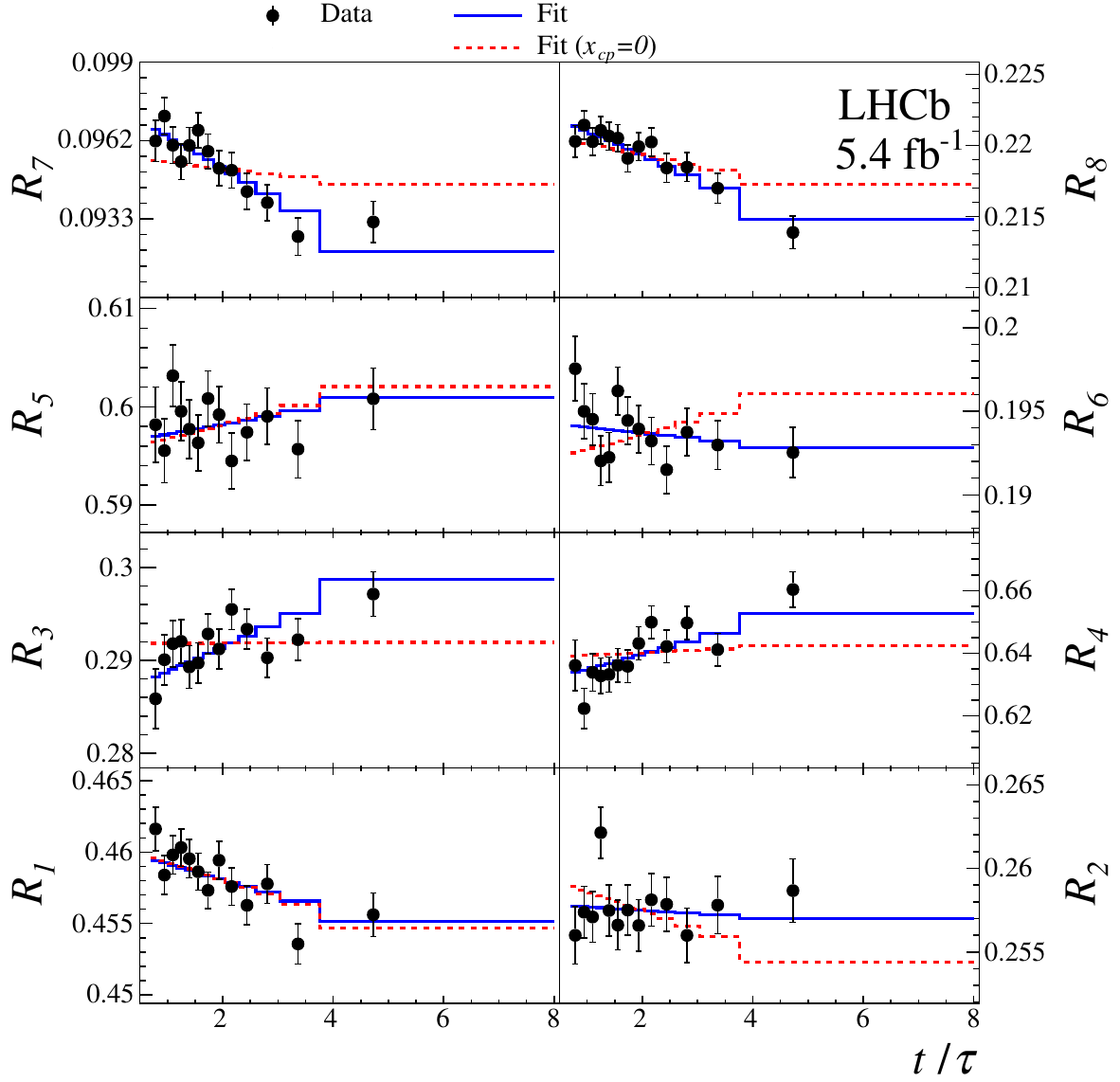}\\
\includegraphics[width=0.7\textwidth]{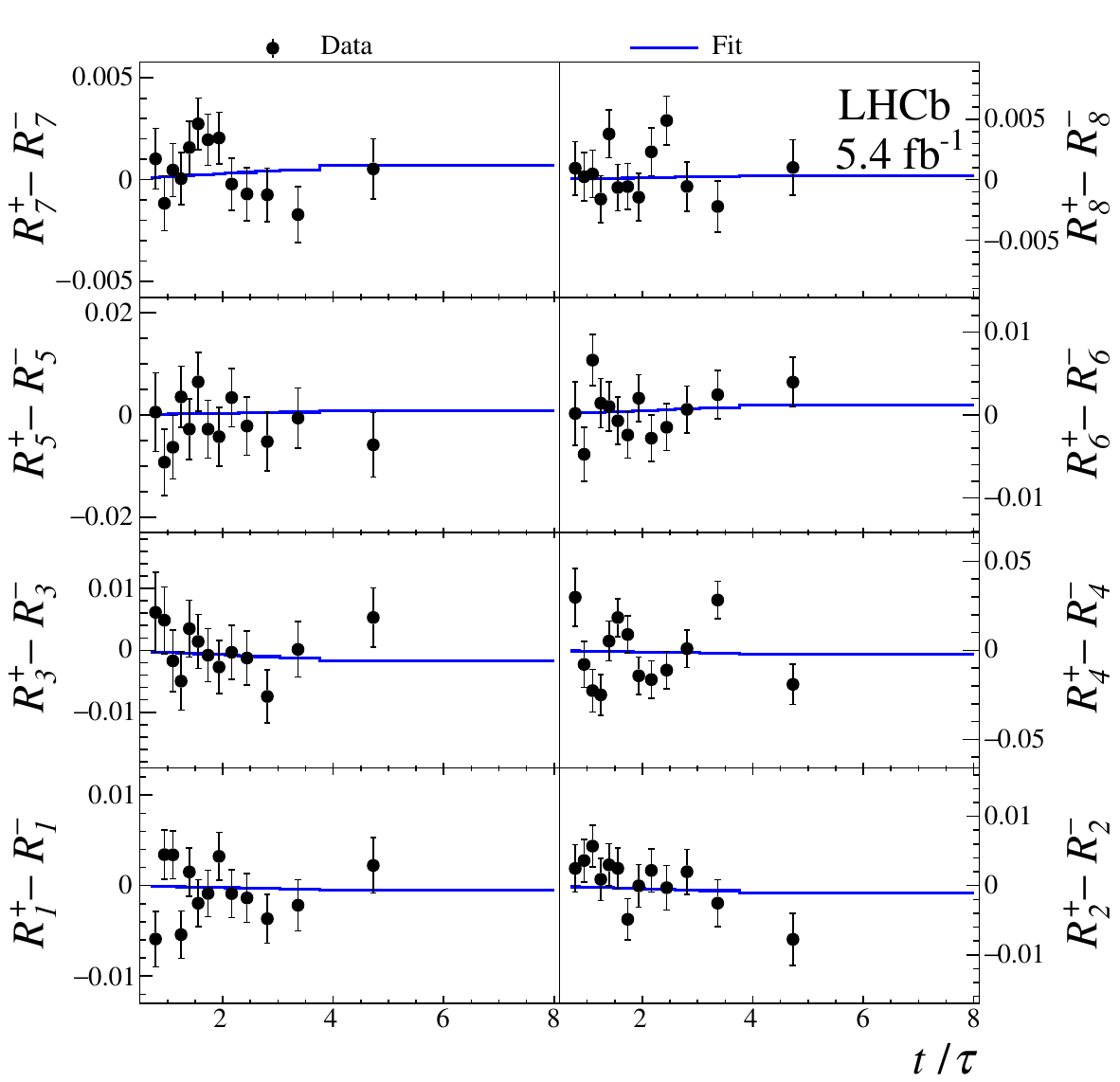}\\
\caption{(Top)
\CP-averaged yield ratios and (bottom) differences of \Dz and \Dzb yield ratios 
as a function of $t/\tau$, shown for each Dalitz-plot bin with fit projections overlaid. \label{fig:fit-results}}
\end{figure}

The mixing parameters are determined by minimizing a least-squares function
\begin{align}
\chisq \equiv& \sum_{+, -}\sum_{b,j}\frac{\left[N^\pm_{-bj}-R^\pm_{+bj}N^\pm_{+bj}/(C_{bj}(1\pm\Delta A_{b}))\right]^2}{(\sigma^\pm_{-bj})^2+\left[R_{+bj}^\pm\sigma^\pm_{+bj}/(C_{bj}(1\pm\Delta A_{b}))\right]^2}\nonumber\\
&+\sum_{b,b'}\left(X^{\rm{EXT}}_b-X_b\right)(V_{\rm{EXT}}^{-1})_{bb'}\left(X^{\rm{EXT}}_{b'}-X_{b'}\right) \label{eq:fit-chi2}\nonumber \\
&+\sum_{b,b'}\left(\Delta A^{\rm{asym}}_b-\Delta A_b\right)(\Delta V_{\rm{asym}}^{-1})_{bb'}\left(\Delta A^{\rm{asym}}_{b'}-\Delta A_{b'}\right),
\end{align}
where the yields $N$ and their measured uncertainties $\sigma$ are scaled by factors for the correlation removal, $C_{bj}$, and detection asymmetry correction, $\Delta A^{b} \equiv \adetpipib - \adetpipinb$. The different subsamples are fitted simultaneously, separated between \Dz and \Dzb flavors denoted as $+$ and $-$, including all decay-time intervals $j$ and Dalitz-plot bins $b$. The parameters $X_b$ are constrained with a Gaussian penalty term using the values $X_b^{\rm{EXT}}$ and covariance matrix $V_{\rm{EXT}}$ from a combination of CLEO and BESIII measurements\cite{Libby:2010nu,Ablikim:2020lpk}. In the fit, the parameters $r_b$ are determined independently for each subsample, as they are affected by the sample-specific variation of the efficiency over the Dalitz plot~\cite{binflip-paper}. To avoid experimenter's bias, the values of \xcp, \ycp, \deltax, and \deltay were not examined until the full procedure had been finalized. Figure \ref{fig:fit-results} shows the yield ratios with fit projections overlaid for each of the eight Dalitz-plot bins. Deviations from constant values are due to mixing. The fit projection when \xcp is fixed to zero is also included and shows the inability of a nonzero \ycp value to produce the deviations on its own. Also shown are the differences of ratios between \Dz and \Dzb decays, where a significant slope would indicate \CP violation.

Systematic uncertainties are assessed from ensembles of pseudoexperiments. These use the \Dkspp model of Ref.~\cite{Adachi:2018itz} to describe the amplitude at $t=0$, and the decay-time dependence is incorporated for a range of values of the mixing and \CP violation parameters. Different sources of systematic uncertainty are included, and the effect on the measured parameters evaluated. The dominant systematic uncertainty on the mixing parameters comes from reconstruction and selection effects, and amounts to $0.20\times10^{-3}$ ($0.76\times10^{-3}$) for $\xcp$ ($\ycp$). This includes neglecting the decay-time and $m_\pm^2$ resolutions and efficiencies, as well as the correction to remove the efficiency correlations. The most important effect for $\ycp$ is the approximation of the strong phase to be constant within each bin in the procedure to remove correlations. Contamination from $b$-hadron decays contributes $0.20\times10^{-3}$ ($0.15\times10^{-3}$) to the $\xcp$ ($\ycp$) uncertainty. Potential mismodeling in the signal yield fits contributes $0.36\times10^{-3}$ to the $\ycp$ uncertainty. Time-dependent detection asymmetries are present mainly in bins that give the best sensitivity to \deltay, resulting in a systematic uncertainty of $0.12\times10^{-3}$.

The consistency of the results is tested by repeating the analysis in subsets of the data, divided according to magnet polarity, trigger and \KS category, data-taking period, \Dstarp meson kinematics, and other categories. The largest variation occurs for the value of \xcp as a function of \Dstarp meson pseudorapidity, where the compatibility, considering statistical uncertainties only, amounts to a p-value of 1.5\%, depending on the details of the sample split, whereas the overall p-value for all \xcp variations observed is above 8\%. The observed variations of the observables \xcp, \ycp, \deltax and \deltay are all consistent with statistical fluctuations.

The mixing and \CP violation parameters are measured to be
\begin{align}
\xcp &= (\phantom{-}\xcpRes\pm\xcpStat\pm\xcpSyst )\times10^{-3},  \nonumber \\
\ycp &= (\phantom{-}\ycpRes\pm\ycpStat\pm\ycpSyst)\times10^{-3}, \nonumber \\
\deltax &= (\dxRes\pm\dxStat\pm\dxSyst)\times10^{-3},  \nonumber \\
\deltay &= (\phantom{-}\dyRes\pm\dyStat\pm\dySyst)\times10^{-3},\nonumber
\end{align}
where the first uncertainty is statistical and the second systematic. The statistical uncertainty contains a subleading component due to the limited precision of the external measurements of the strong phases and control samples used for the detection asymmetry. This amounts to approximately $(0.23$, $0.66$, $0.04$, and $0.08)\times 10^{-3}$ for \xcp, \ycp, \deltax, and \deltay, respectively. The measurements are statistically limited, though the systematic uncertainty on \ycp is comparable to the statistical uncertainty. The results are used to form a likelihood function of $x$, $y$, $|q/p|$, and $\phi$ using a likelihood-ratio ordering that assumes the observed correlations to be independent of the true parameter values~\cite{LHCb-PAPER-2013-020}. The best fit point is 
\begin{align}
 x &= (3.98_{-\,0.54}^{+\,0.56})\times10^{-3}, \nonumber \\
 y &= (\phantom{0}4.6_{-\,1.4}^{+\,1.5\phantom{0}})\times10^{-3}, \nonumber \\
 |q/p| &= 0.996\pm0.052, \nonumber \\
 \phi &= -0.056_{-\,0.051}^{+\,0.047}. \nonumber
\end{align}

In summary, a measurement of mixing and \CP violation in $\Dkspp$ decays has been performed with the bin-flip method, using $pp$ collision data collected by the \lhcb experiment and corresponding to an integrated luminosity of 5.4\invfb. This resulted in the first observation of a nonzero value of the mass difference $x$ of neutral charm meson mass eigenstates with a significance of more than seven standard deviations, and significantly improves limits on mixing-induced \CP violation in the charm sector.

\section*{Acknowledgements}
%
%
\noindent We express our gratitude to our colleagues in the CERN
accelerator departments for the excellent performance of the LHC. We
thank the technical and administrative staff at the LHCb
institutes.
We acknowledge support from CERN and from the national agencies:
CAPES, CNPq, FAPERJ and FINEP (Brazil); 
MOST and NSFC (China); 
CNRS/IN2P3 (France); 
BMBF, DFG and MPG (Germany); 
INFN (Italy); 
NWO (Netherlands); 
MNiSW and NCN (Poland); 
MEN/IFA (Romania); 
MSHE (Russia); 
MICINN (Spain); 
SNSF and SER (Switzerland); 
NASU (Ukraine); 
STFC (United Kingdom); 
DOE NP and NSF (USA).
We acknowledge the computing resources that are provided by CERN, IN2P3
(France), KIT and DESY (Germany), INFN (Italy), SURF (Netherlands),
PIC (Spain), GridPP (United Kingdom), RRCKI and Yandex
LLC (Russia), CSCS (Switzerland), IFIN-HH (Romania), CBPF (Brazil),
PL-GRID (Poland) and NERSC (USA).
We are indebted to the communities behind the multiple open-source
software packages on which we depend.
Individual groups or members have received support from
ARC and ARDC (Australia);
AvH Foundation (Germany);
EPLANET, Marie Sk\l{}odowska-Curie Actions and ERC (European Union);
A*MIDEX, ANR, IPhU and Labex P2IO, and R\'{e}gion Auvergne-Rh\^{o}ne-Alpes (France);
Key Research Program of Frontier Sciences of CAS, CAS PIFI, CAS CCEPP, 
Fundamental Research Funds for the Central Universities, 
and Sci. \& Tech. Program of Guangzhou (China);
RFBR, RSF and Yandex LLC (Russia);
GVA, XuntaGal and GENCAT (Spain);
the Leverhulme Trust, the Royal Society
 and UKRI (United Kingdom).

\addcontentsline{toc}{section}{References}
\bibliographystyle{LHCb}
\bibliography{main,standard,LHCb-PAPER,LHCb-CONF,LHCb-DP,LHCb-TDR}

\clearpage
\section{Supplemental material\label{supplemental-material}}
\label{sec:Supplemental}

\begin{figure}[h!!]
\centering
\includegraphics[width=0.5\textwidth]{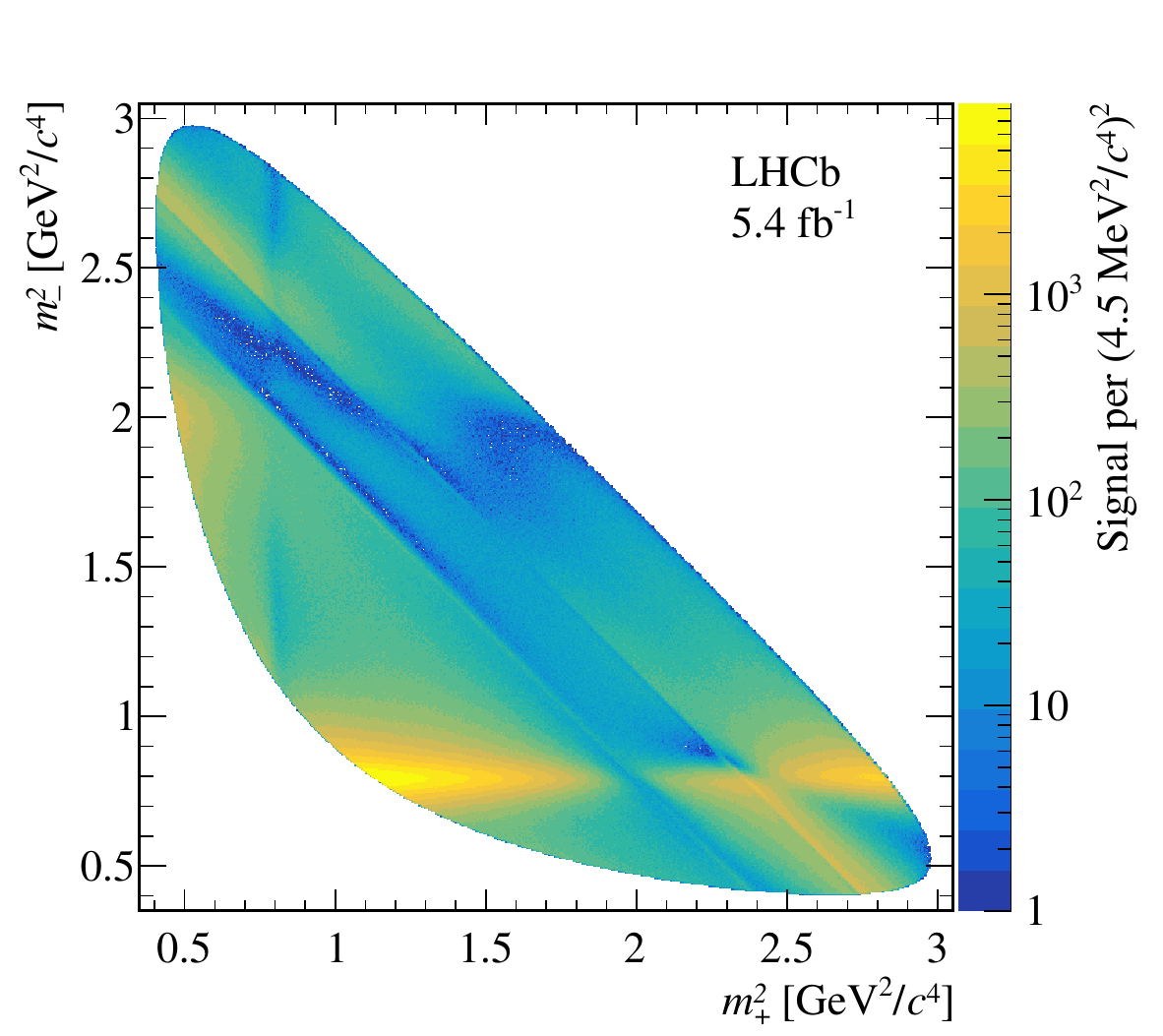}\hfil
\caption{Dalitz plot of background-subtracted \Dkspp candidates. \label{fig:dalitz}}
\end{figure}

\cref{tab:fit-results} summarizes the measured values along with their uncertainties and correlations. \cref{tab:derived-results} gives the derived values for $x$, $y$, $q/p$ and $\phi$ together with the 95.5\% confidence interval. \cref{tab:uncertainties} shows a summary of the uncertainties in this analysis. \cref{fig:dalitz} shows the Dalitz plot of the background-subtracted \Dkspp candidates used in the analysis. No efficiency corrections are applied. All samples are combined. 

\begin{table}[h!]
\centering
\caption{Fit results of \xcp, \ycp, \deltax, and \deltay. The first contribution to the uncertainty is statistical, the second systematic.}\label{tab:fit-results}
\begin{tabular}{lcrrrrrr}
\toprule
Parameter & Value & \multicolumn{3}{c}{Stat.\ correlations} & \multicolumn{3}{c}{Syst.\ correlations}\\
 & [$10^{-3}$] & \ycp & \deltax & \deltay & \ycp & \deltax & \deltay\\
\midrule
\xcp & $\phantom{-}\xcpRes\pm\xcpStat\pm\xcpSyst$  & $0.11$ &  $-0.02$ & $-0.01$ &  $0.13$ &  $0.01$ & $0.01$\\
\ycp & $\phantom{-}\ycpRes\pm\ycpStat\pm\ycpSyst$  &         & $-0.01$ &  $-0.05$ &         & $-0.02$ & $0.01$\\
\deltax & $\dxRes\pm\dxStat\pm\dxSyst$  &         &         & $0.08$ &         &         &  $0.31$\\
\deltay & $\phantom{-}\dyRes\pm\dyStat\pm\dySyst$  &         &         & \\
\bottomrule
\end{tabular}
\end{table}

\begin{table}[t]
\centering
\caption{Point estimates and 95.5\% confidence-level (CL) intervals for $x$, $y$, $|q/p|$ and $\phi$. The uncertainties include statistical and systematic contributions.\label{tab:derived-results}}
\begin{tabular}{ccc}
\toprule
Parameter & Value & 95.5\% CL interval \\
\midrule
$x$ [$10^{-3}$] & $\phantom{-0}3.98\,_{-\,0.54}^{+\,0.56\phantom{0}}$ & $\phantom{-00}[2.9,5.0]\phantom{00}$ \\
$y$ [$10^{-3}$] & $\phantom{-00}4.6\,_{-\,1.4}^{+\,1.5\phantom{00}}$ & $\phantom{-00}[2.0,7.5]\phantom{00}$ \\
$|q/p|$         & $\phantom{-00}0.996\pm0.052$ & $[\phantom{-}0.890,1.110]$ \\
$\phi$          & $-0.056\,_{-\,0.051}^{+\,0.047}$ & $[-0.172,0.040]$ \\
\bottomrule
\end{tabular}
\end{table}

\begin{table}[h]
\centering
\caption{Uncertainties in units of $10^{-3}$. The total systematic uncertainty is the sum in quadrature of the individual components. The uncertainties due to the external inputs and detection asymmetry calibration samples are included in the statistical uncertainty. These are also reported separately, along with the contributions due to the limited sample size, to ease comparison with other sources. \label{tab:uncertainties}}
\begin{tabular}{lcccc}
\toprule
Source & \xcp & \ycp & \deltax & \deltay \\
\midrule
 Reconstruction and selection  & $0.199 $ & $0.757 $ & $0.009 $ & $0.044 $\\
 Secondary charm decays  & $0.208 $ & $0.154 $ & $0.001 $ & $0.002 $\\
 Detection asymmetry  & $0.000 $ & $0.001 $ & $0.004 $ & $0.102 $\\
 Mass-fit model  & $0.045 $ & $0.361 $ & $0.003 $ & $0.009 $\\
\midrule
 Total systematic uncertainty  & $0.291 $ & $0.852 $ & $0.010 $ & $0.110 $\\
\bottomrule
\\
\toprule
Strong phase inputs& 0.23 & 0.66 & 0.02 & 0.04 \\
Detection asymmetry inputs& 0.00 & 0.00 & 0.04 & 0.08 \\
Statistical (w/o inputs)& 0.40 & 1.00 & 0.18 & 0.35 \\
\midrule
Total statistical uncertainty & \xcpStat  & \ycpStat  & \dxStat  & \dyStat  \\
\bottomrule
\end{tabular}
\end{table}

\clearpage



\newpage
\centerline
{\large\bf LHCb collaboration}
\begin
{flushleft}
\small
R.~Aaij$^{32}$,
C.~Abell{\'a}n~Beteta$^{50}$,
T.~Ackernley$^{60}$,
B.~Adeva$^{46}$,
M.~Adinolfi$^{54}$,
H.~Afsharnia$^{9}$,
C.A.~Aidala$^{86}$,
S.~Aiola$^{25}$,
Z.~Ajaltouni$^{9}$,
S.~Akar$^{65}$
% ...

{\footnotesize \it

% ...
$^{9}$Universit{\'e} Clermont Auvergne, CNRS/IN2P3, LPC, Clermont-Ferrand, France\\
% ...
$^{25}$INFN Sezione di Milano, Milano, Italy\\
% ...
$^{32}$Nikhef National Institute for Subatomic Physics, Amsterdam, Netherlands\\
% ...
$^{46}$Instituto Galego de F{\'\i}sica de Altas Enerx{\'\i}as (IGFAE), Universidade de Santiago de Compostela, Santiago de Compostela, Spain\\
% ...
$^{50}$Physik-Institut, Universit{\"a}t Z{\"u}rich, Z{\"u}rich, Switzerland\\
% ...
$^{54}$H.H. Wills Physics Laboratory, University of Bristol, Bristol, United Kingdom\\
% ...
$^{60}$Oliver Lodge Laboratory, University of Liverpool, Liverpool, United Kingdom\\
% ...
$^{65}$University of Cincinnati, Cincinnati, OH, United States\\
% ...
$^{68}$Syracuse University, Syracuse, NY, United States\\
% ...
$^{86}$University of Michigan, Ann Arbor, United States, associated to $^{68}$\\
\bigskip
% ...
\medskip
}
\end{flushleft}

\end{document}